\documentclass[twocolumn,aps,prc,superscriptaddress,nofootinbib,showpacs,floatfix]{revtex4-1}
\usepackage{url}
\usepackage{cancel}
\usepackage[colorlinks,linkcolor=blue,citecolor=blue,filecolor=black,urlcolor=blue]{hyperref}
\usepackage{epsfig,graphics}
\usepackage{graphicx}
\usepackage{dcolumn}
\usepackage{bm}
\usepackage[usenames]{color}
\usepackage{amssymb}
\usepackage{amsmath}
\usepackage{multirow}
\usepackage{enumitem}
\usepackage{float}
\usepackage{harpoon}
\usepackage{MnSymbol}
\usepackage{appendix}
\usepackage{color}
\makeatletter

\newcommand{\Rmnum}[1]{\expandafter\@slowromancap\romannumeral #1@}
\makeatother

\makeatletter 
\@addtoreset{equation}{section}
\makeatother  

\newcommand {\roots}{\mbox{$\sqrt{s}$}}
\newcommand {\rootsmp}{\mbox{$\sqrt{s_{\mu p}}$}}

\usepackage{CJK}
\hfuzz=\maxdimen
\begin{document}
\title{A Muon-Ion Collider at BNL: the future QCD frontier 
and path to a new energy frontier of $\mu^+\mu^-$
colliders}
\newcommand{\florida}{Department of Physics, 
University of Florida, Gainesville, Florida 32611, USA}
\newcommand{\rice}{Physics Department, Rice University, Houston, Texas 77251, USA}
 \author{Darin Acosta}\email{dea6@rice.edu}\affiliation{\rice}
 \author{Wei Li}\email{wl33@rice.edu}\affiliation{\rice}

\begin{abstract}
We propose the development and construction of a novel muon-ion 
collider (MuIC) at Brookhaven National Laboratory (BNL) in the USA 
as an upgrade to succeed the electron-ion collider (EIC) that 
is scheduled to commence in the early 2030s, by a joint effort of 
the nuclear and particle physics communities. The BNL facility could
accommodate a muon storage beam with an energy up to about 1~TeV 
with existing magnet technology. When collided with a 275~GeV hadron
beam, the MuIC center-of-mass energy of about 1~TeV will extend 
the kinematic coverage of deep inelastic scattering physics 
at the EIC (with polarized beams) by more than 
an order of magnitude in $Q^2$ and $x$, opening a 
new QCD frontier to address many fundamental
scientific questions in nuclear and particle physics. 
This coverage is comparable to that of the proposed Large
Hadron-Electron Collider (LHeC) at CERN, but with complementary lepton 
and hadron kinematics, ion species and beam polarization. 
Additionally, the development
of a MuIC at BNL will focus the worldwide R\&D efforts on muon 
collider technology and serve as a demonstrator toward a future
muon-antimuon collider at {$\mathcal O$}(10)~TeV energies, which is an 
attractive option to reach the next high energy frontier 
in particle physics at an affordable cost and a smaller 
footprint than a future circular hadron collider. 
We discuss here the possible design parameters of the MuIC, 
kinematic coverage, science cases, and detector design 
considerations including estimates of resolutions on DIS 
kinematic variables. A possible road map toward the future 
MuIC and muon-antimuon colliders is also presented.\end{abstract}
             
\maketitle

\section{Introduction}

Lepton-hadron (nucleus) deep inelastic scattering (DIS) 
has been a powerful tool to understand the fundamental 
structure of nucleons and nuclei. Decades of DIS
experiments have revealed the point-like substructure of
quarks and gluons inside the nucleon, and how they share
the longitudinal momentum of a fast-moving nucleon.
To develop a deeper understanding of the quark-gluon
structure and dynamics (especially in three dimensions) 
of matter, governed by
quantum chromodynamics (QCD), a high energy and 
luminosity polarized electron-ion collider (EIC) has recently
been endorsed to be built at Brookhaven National Laboratory 
(BNL) by the late 2020s~\cite{NAP25171} as a high priority
on the agenda of the US nuclear physics community. The EIC 
is capable of carrying out deep inelastic electron-proton 
and electron-nucleus collisions with polarized beams 
at a center-of-mass energy ($\roots$) up to
140~GeV~\cite{Accardi:2012qut,NAP25171}. It will 
establish a new QCD frontier to address key open questions
such as the origin of nucleon spin, mass, and the emergence of QCD
many-body phenomena at extreme parton densities. At CERN, 
the Large Hadron-electron Collider
(LHeC)~\cite{Agostini:2020fmq} has been proposed as a possible 
extension to the Large Hadron Collider (LHC)
to explore even higher energy regimes of $\roots \approx 1$~TeV 
and beyond. As a potential long-term step at CERN, 
the Future Circular Collider (FCC) proposed to be built in a new 100~km tunnel 
close to CERN also includes a mode of electron-hadron collisions 
(FCC-he) at $\roots$ = 3.5~TeV~\cite{Abada:2019lih}. 

We propose an alternative approach to achieve the next-generation
lepton-hadron (ion) collider at the 1-TeV scale using high-energy
muon beams, a Muon-Ion Collider (MuIC), based on existing facilities
at BNL after the mission of EIC is completed by late-2030s. 
Possibilities of muon-hadron colliders and their scientific 
potential have been discussed previously, 
for example in Refs.~\cite{Shiltsev:1997pv,Ginzburg:1998yw,Sultansoy:1999na,Acar:2016rde,Canbay:2017rbg,Ketenoglu:2018fai,Kaya:2019ecf,Cheung:2021iev}. 
A MuIC at BNL would serve as a major 
step toward a future muon-antimuon ($\mu^{+}\mu^{-}$) collider, 
which has received revived attention in the particle 
physics community in recent years because of 
its potential of reaching $\roots = {\mathcal O}$(10)~TeV 
in a relatively compact tunnel (e.g., the size of the LHC). It has 
been argued that for the production cross 
section of hard processes or heavy unknown particles, 
the discovery potential of this
particle at a 14~TeV $\mu^{+}\mu^{-}$ collider 
matches that of a 100~TeV proton-proton
collider (e.g., FCC)~\cite{Delahaye:2019omf}, 
as all the available beam energy is carried by the interacting muons. 
While it is in principle feasible 
to build the next electron/proton colliders based on established 
technologies and advanced magnets with the construction of a new, 
longer {$\mathcal O$}(100)~km tunnel, 
affordability is a concern in terms of both 
cost and time. For this reason, exploring new accelerator 
and collider approaches provides attractive alternatives.
The MuIC and $\mu^{+}\mu^{-}$ collider also share
synergies with the proposed ``Neutrinos from STORed Muons''
facility~\cite{Adey:2015iha}.

Development of muon collider technology is still at
an early stage, with many challenges to overcome (e.g., see details in Ref.~\cite{Delahaye:2019omf}). For example,
because of the short lifetime of the muon, even accelerated to TeV 
energies, the beam will constantly decay, and thus will require
rapid acceleration and collisions to maintain sufficiently high 
luminosity. To achieve a high luminosity, cooling of muon bunches 
to reduce its phase space before acceleration is
crucial~\cite{MICE:2019jkl}.
Beam backgrounds from muon decays also pose challenges  
to both the accelerator and the detectors. 
Radiation hazards from interacting neutrinos would need 
to be mitigated, especially at {$\mathcal O$}(10)~TeV energies. 
There is a consensus in the muon collider community that 
realizing a smaller-scale muon 
collider would be a necessary intermediate step to serve 
as a demonstrator before pursing the ultimate
{$\mathcal O$}(10)~TeV $\mu^{+}\mu^{-}$ collider. Such a demonstrator 
still requires tremendous R\&D efforts and significant cost, so 
a compelling science program that is not
accessible by other proposed facilities is needed. For example,
the physics of a $\mu^{+}\mu^{-}$ collider with $\roots$ of 
several hundred GeV to 1 TeV may not be competitive with 
an $e^+e^-$ collider proposed with more 
established technology, such as the International
Linear Collider (ILC)~\cite{Behnke:2013xla}, the Compact Linear
Collider (CLIC)~\cite{Charles:2018vfv}, the Circular Electron Positron
Collider (CEPC)~\cite{CEPCStudyGroup:2018ghi}, and 
the FCC-ee~\cite{Abada:2019lih}.

In this paper, we argue that a Muon-Ion Collider (MuIC) at BNL
will fulfill the objectives of both establishing a new QCD
frontier to succeed the EIC, and serve as 
a demonstrator of muon collider technology toward 
an ultimate {$\mathcal O$}(10)~TeV $\mu^{+}\mu^{-}$ collider.
Such an intermediate machine will be of great interest to both nuclear and 
particle physics communities worldwide. Re-using the existing
facility and infrastructure is a key to make the project
much more affordable. In the following sections, 
we discuss potential key parameters and designs of the proposed MuIC, 
including rough estimates of achievable luminosity in 
Section~\ref{sec:muic}, its scientific potential in 
Section~\ref{sec:science}, the final-state kinematics 
in Section~\ref{sec:kinematics}, and
detector design considerations in Section~\ref{sec:detector}. 
Finally, a possible road map toward realizing the MuIC with a joint
effort of the nuclear and particle physics communities is
discussed in Section~\ref{sec:roadmap}.

\section{A Muon-Ion Collider at BNL}
\label{sec:muic}

A key merit of our proposal is to re-use the existing
facility and infrastructure to open new avenues and 
advance new technologies in both nuclear and particle physics 
at a reasonable cost. Some key parameters
of the proposed muon-ion collider based at BNL are listed in
Tables~\ref{tab:table} and ~\ref{tab:table2}, 
taking muon-proton and muon-Au collisions as examples. 
These parameters were chosen based on proposed $\mu^{+}\mu^{-}$
colliders~\cite{Delahaye:2019omf} and the EIC being planned 
at BNL~\cite{Aschenauer:2014cki}.

\subsection{Design Concept}

The EIC will be constructed based on the existing Relativistic
Heavy Ion Collider (RHIC). The RHIC tunnel is built in a hexagon 
with rounded corners. A new linac following by the 
Rapid Cycling Synchrotron (RCS) will be added in the RHIC 
tunnel for electron injection and acceleration. A new 
electron storage ring will be added or be converted 
from one of the two existing hadron rings 
for collision. To evolve the EIC into a MuIC, linac+RCS is 
among possible proposed technologies for muon
acceleration so could be reused. 
The electron storage ring would be converted into that for 
muon beams by replacing with corresponding magnets 
(see discussions below). 
It is anticipated that additional shieldings are 
needed to protect against backgrounds from muon 
decays. To generate high intensity muon sources, 
there are two proposed approaches for the muon 
collider technology: the proton driver scheme 
followed by the muon cooling and the positron 
driver scheme (no cooling is needed)~\cite{Delahaye:2019omf}.
The proton driver scheme is presumably more suitable 
for BNL, taking advantage of existing high intensity 
proton sources.

In Table~\ref{tab:table}, we discuss possible
energy scenarios of the MuIC at BNL. Assuming no upgrades to the
hadron beam, the maximum proton beam energy 
at the EIC is 275~GeV. The achievable muon beam
energy can be estimated by $p^{\mu}=0.3Br$, where $r$
is the bending radius and $B$ is the bending magnetic 
field of the storage ring.
The bending radius is 290~m for the electron storage 
ring at the EIC. We consider three possible scenarios 
of the dipole bending magnets for the muons: 8.4~T 
currently used at the LHC, 11~T being developed for the 
HL-LHC~\cite{Bordini:2019fan}, and 16~T aimed for the 
future FCC~\cite{Benedikt:2018csr} requiring significant R\&D 
effort. All three scenarios provide a muon-nucleon 
center-of-mass energy ($\rootsmp$) of about 1~TeV
(within $\pm$12\%), a 10-fold increase to the EIC energy. 
A conservative approach would be to use the 8.4~T LHC
dipole magnets, which give a muon beam energy of 730~GeV.
Even considering a more conservative scenario of 3.8~T 
magnets being used by the RHIC hadron ring would still lead 
to $\rootsmp$ that is several times that of the EIC.
In the discussion below, we will focus on a scenario of 
11~T bending magnets planned for the HL-LHC, 
which is realistically achievable and 
can accommodate a muon beam energy of 960~GeV. In this scenario,
the maximum achievable $\rootsmp$ for muon-Au collisions
is 650~GeV with $^{197}$Au ion beam energy of 110~GeV 
per nucleon.

\begin{table}[t!]
\centering
\caption{The proposed energies of the MuIC at BNL for three possible magnet scenarios.}
\vspace{0.5cm}
\includegraphics[width=\linewidth]{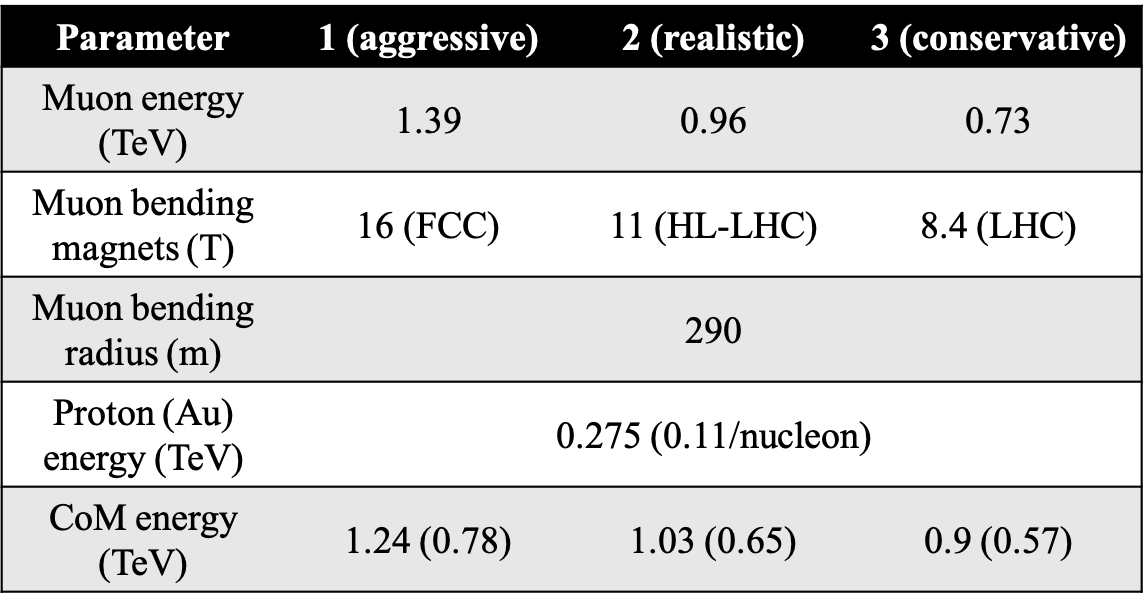}
\label{tab:table}
\end{table}

\subsection{Luminosity Estimates}

The instantaneous luminosity of a muon-proton collider
is discussed in Ref.~\cite{Kaya:2019ecf} and 
can be generally expressed as follows:

\begin{equation}
    \mathcal{L}_{\mu p} = \frac{N^{\mu}N^{p}}{4\pi\max[\sigma^{\mu}_{x},\sigma^{p}_{x}]\max[\sigma^{\mu}_{y},\sigma^{p}_{y}]}\min[f^{\mu}_{c},f^{p}_{c}]H_{hg},
\end{equation}

\noindent Here, $N^{\mu}$ and $N^{p}$ represent the number of
particles per respective beam bunch.  
The transverse
RMS beam size in $x$ and $y$ for the muon and proton
beam is $\sigma^{\mu,p}_{x,y}$, and $f^{\mu,p}_{c}$ is the bunch frequency. Typically, 
the bunch frequency of proton beams is 2--3 orders of magnitude
larger than that of muon beams, so ${L}_{\mu p}$ is largely
determined by $f^{\mu}_{c}$, which is equal to the muon bunch 
repetition frequency ($f_{\rm rep}$) multiplied by the number of
cycles ($N_{c}$) muons can make in a circular storage ring 
before decaying away. For simplicity, the hour-glass factor, 
$H_{hg}$, is assumed to be unity.

\begin{table}[t!]
\centering
\vspace{0.5cm}
\includegraphics[width=0.95\linewidth]{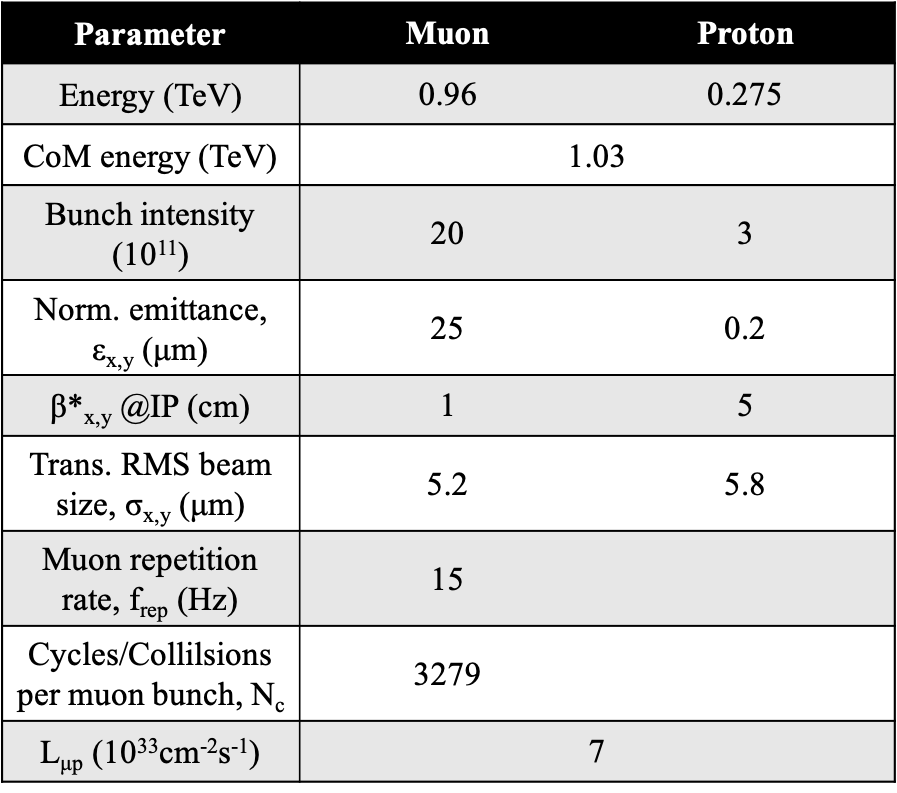}
\caption{Rough estimates of achievable luminosity for the
proposed MuIC at BNL with the muon beam energy of 960~GeV.}
\label{tab:table2}
\end{table}

We use proposed 
parameters of the proton driver scheme for our 
estimate of MuIC luminosity below. As listed in
Table~\ref{tab:table2}, the muon bunch repetition 
frequency is taken to be 15~Hz, and the transverse RMS beam size,
$\sigma_{x,y}=\sqrt{\beta_{x,y}\varepsilon/\gamma}$,
is estimated to be about 5--6~$\mu$m for the muon 
beam based on proposed muon 
colliders~\cite{Palmer:2014nza,Delahaye:2019omf}
and the proton beam achieved at
RHIC~\cite{Aschenauer:2014cki}. Each muon 
bunch will survive an average of about 300B(Tesla) cycles 
in a ring. These parameters lead to a $\mathcal{L}_{\mu p}$ of
5--10$\times$10$^{33}$~cm$^{-2}$s$^{-1}$, comparable 
to the LHeC design luminosity.
For simplicity, the estimated MuIC luminosity 
in this paper assumes a round transverse 
beam profile for both the muon and proton beams.
However, note that the EIC design 
at BNL adopts a flat transverse beam profile with the horizontal
dimension stretched much larger than the vertical 
dimension. At the interaction point, the two beams 
would intersect at a finite crossing angle of 25~mrad 
in the horizontal plane. The purpose of such a design is to 
maximize the luminosity and, at the same time, fulfill 
other requirements such as the possibility of detecting scattered 
protons with a transverse momentum as low as 200~MeV by Roman Pots 
inside the beam pipe. 
Since there may still be large uncertainties in 
muon collider parameters, the  
values of luminosity presented should be taken as an order of 
magnitude estimate, likely optimistic. More sophisticated 
considerations would need 
to be taken into account for a practical MuIC design.

To put the MuIC luminosity into context, it is anticipated that the EIC 
will deliver an integrated luminosity up to about 1.5~fb$^{-1}$/month with
$\mathcal{L}_{ep} \approx 10^{33}$~cm$^{-2}$s$^{-1}$,
and most of science cases studied at the EIC require a total
integrated luminosity of 10~fb$^{-1}$ ($\approx 30$ weeks of operations). 
Therefore, even with much less stringent requirements on the 
muon beam, e.g., a peak $\mathcal{L}_{\mu p} \approx  10^{32}$~cm$^{-2}$s$^{-1}$ operating for several years, 
the MuIC will still be a novel facility that breaks new ground 
in science and technology of high energy nuclear and particle 
physics. Generally speaking, higher luminosity will help 
access physics at the high $Q^{2}$ and electroweak regimes.

\subsection{Beam polarization}
Both the electron and hadron beams at the EIC are designed 
to be highly polarized ($\approx 70$\%), a unique capability for 
studying the origin of the nucleon's spin. 
Muons produced from pion decays are naturally
polarized. The level of polarization in the lab frame depends
on the initial pion energy and decay angle. The average natural
polarization of captured muons is about
20\%~\cite{Norum:1996mi,Neuffer:1999aw,Ankenbrandt:1999cta}.
It is possible to achieve higher polarization by, e.g., 
selecting on the energy of forward-going muons at the cost
of luminosity. Studies have suggested that about 50\% 
polarization can be achieved at a reduced luminosity of about
25\% of the maximum value~\cite{Neuffer:1999aw,Ankenbrandt:1999cta}. 
Therefore, the MuIC will retain the unique capability of beam 
polarization at BNL, which is particularly important for the
study of nucleon spin. To what extent the muon polarization can be
maintained through the cooling and acceleration stages is a
subject that requires future studies.
It is suggested in Ref.~\cite{Cline:1996yi} that depolarization 
in the ionization cooling element should be negligible.

\begin{figure}[t!]
\centering
\includegraphics[width=\linewidth]{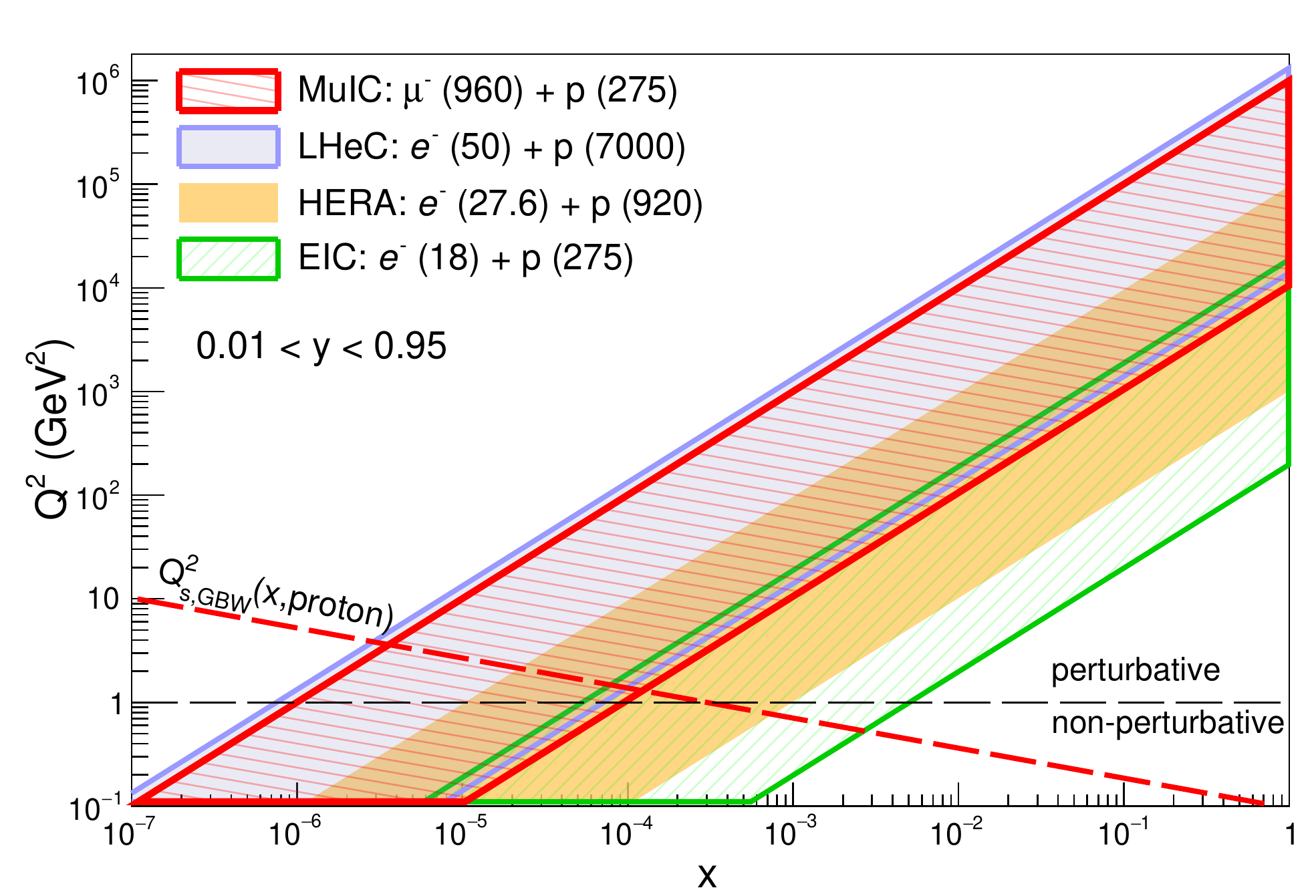}
\includegraphics[width=\linewidth]{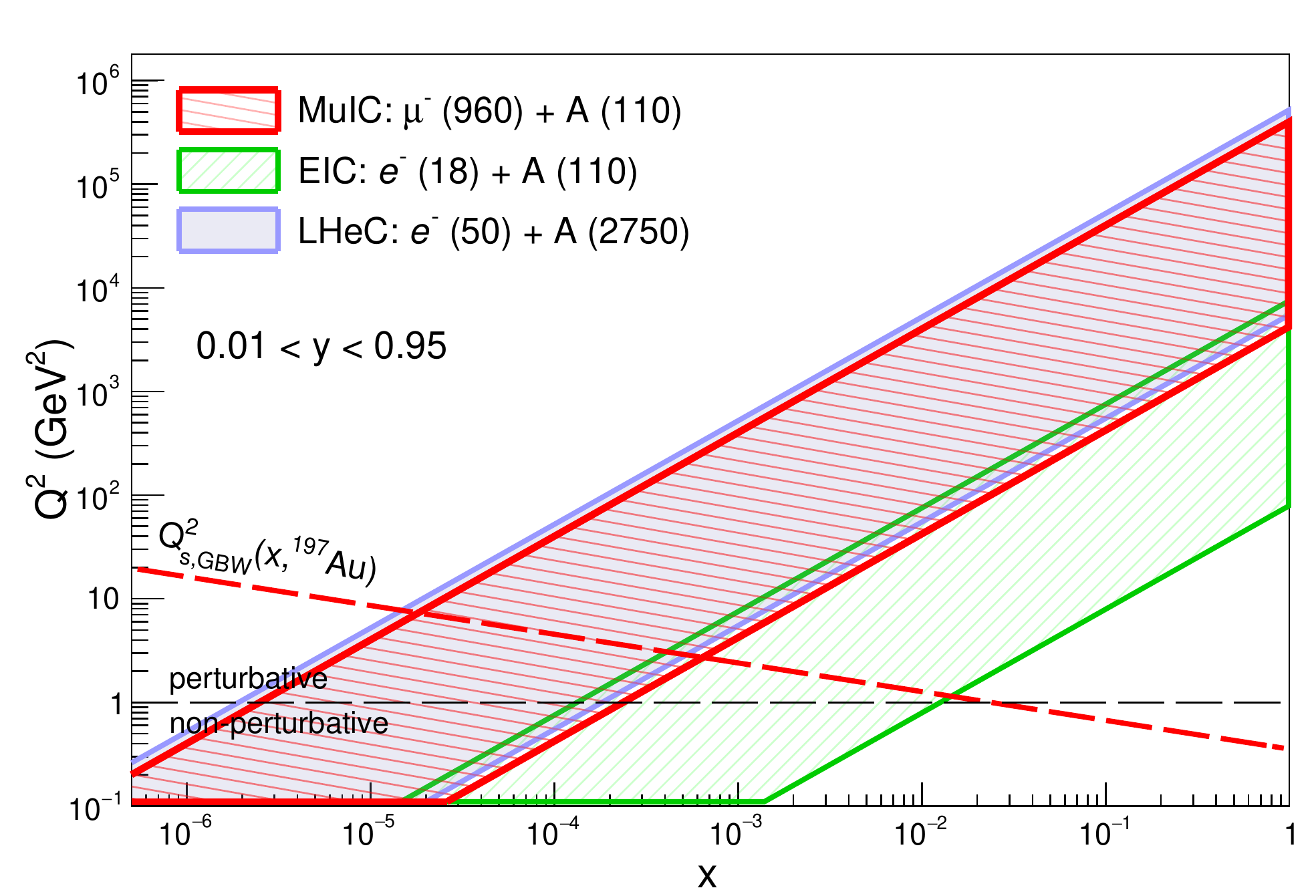}
\caption{Kinematic coverage of $Q^{2}$ and $x$ in deep 
inelastic lepton-proton (top) and lepton-nucleus (bottom)
scattering. The four cases shown are for the EIC at BNL, HERA at DESY, 
and LHeC at CERN, each at their  maximum beam energies, and the proposed 
MuIC at BNL with 960~GeV muon and 275 (110)~GeV proton (ion) 
beams. The inelasticity ($y$) range is assumed to be $0.01<y<0.95$
(hatched areas). The long dashed lines indicate the saturation 
scale as a function of $x$ in the proton and the gold ($^{197}$Au) 
nucleus from the GBW model~\cite{GolecBiernat:1998js}.}
\label{fig1}
\end{figure}

\subsection{Synergy with the nuSTORM facility}
\label{sec:nustorm}

The nuSTORM facility is proposed as a concept
of neutrino factory using a racetrack-like muon storage ring
to deliver high-intensity beams of $\nu_e$ and $\nu_\mu$ from muon decays.
It will enable searches for light sterile neutrinos with 
unprecedented sensitivity over a wide parameter space, and can also
be exploited to make detailed studies of neutrino–nucleus 
scattering. Besides a rich program of nuclear and particle physics, 
the nuSTORM facility also provides a testing ground for developing
high-intensity muon storage ring to facilitate R\&Ds of a muon collider.
See a review in Ref.~\cite{Adey:2015iha}.
Given synergies from several communities, such a facility can also 
be hosted at BNL as part of a testing facility toward the proposed MuIC,
while having a strong physics program on its own. As the
RHIC tunnel consists of six long straight sections, the MuIC with
stored muons of $\sim$1~TeV will provide high flux, high energy
neutrino beams to enable a compelling neutrino physics program
at the same time.

\section{The Science of a Muon-Ion Collider}
\label{sec:science}

To put potential scientific impacts of the proposed MuIC 
in perspective, Fig.~\ref{fig1} compares the kinematic 
coverage of the MuIC in the map of momentum transfer, $Q^{2}$, 
and Bjorken scaling variable, $x$, with the EIC at BNL, 
HERA at DESY and the proposed LHeC at CERN, for lepton-proton (top) and 
lepton-nucleus (bottom) collisions. The boundaries of the $Q^{2}$-$x$ 
coverage are determined by $Q^{2}=sxy$, where $s$ is the 
squared center-of-mass energy and $y$ is the inelasticity. 
Here, we take an inelasticity range of $0.01<y<0.95$ 
that is commonly assumed for $ep$ collisions
(although it generally depends on the detector technology 
and method of reconstructing the DIS kinematics). The 
dashed line at $Q^{2}=1$~GeV$^{2}$ can be considered as 
the transition from the non-perturbative to perturbative 
regime of QCD, which is usually a lower limit required in 
DIS studies.

The red dashed lines in Fig.~\ref{fig1} 
indicate the gluon saturation scales inferred
from the GBW model based on fits to HERA data~\cite{GolecBiernat:1998js},
below which nonlinear QCD evolution effects are expected 
to become significant and the growth of the gluon density toward 
small $x$ values will saturate~\cite{Gribov:1983ivg}. 
Other models on estimating the saturation scale 
can be found in
Refs.~\cite{Albacete:2009fh,Albacete:2010sy,Kowalski:2003hm},
including those that consider the impact parameter dependence.
The saturation scale is predicted 
to be enhanced by a factor $A^{1/3}$ ($\approx 6$ for $^{197}$Au)
in nuclei, where $A$ is the atomic number, because of 
overlapping nucleon gluon fields induced by the Lorentz contraction in the longitudinal
direction~\cite{McLerran:1993ni,Gelis:2010nm}.

As shown by Fig.~\ref{fig1}, both the LHeC and MuIC will significantly
extend the kinematic coverage of the EIC to much larger $Q^{2}$ and
smaller $x$ regimes, by an order of magnitude in each compared to the previous HERA ep collider. In lepton-proton collisions (Fig.~\ref{fig1}, top), 
the saturation regime is clearly out of the reach for the EIC, 
but it becomes within reach at very small $x$ values at the LHeC and MuIC. 
In lepton-nucleus collisions (Fig.~\ref{fig1}, bottom), considering the
$^{197}$Au nucleus as a representative example (with a factor of 6 enhancement in the 
saturation scale), the EIC starts approaching the saturation regime, 
while the LHeC and MuIC should be well in the domain to explore gluon saturation 
and nonlinear QCD phenomena. Because of large overlaps of the 
proposed MuIC and LHeC in terms of the $Q^{2}$-$x$ coverage (although the
kinematics of the final-state particles are quite different in the lab frame, 
as will be discussed later), they share a lot in common for their physics 
potential. The science program of the LHeC has been well presented 
in Ref.~\cite{Agostini:2020fmq}. In addition, the MuIC at BNL would offer the 
unique advantage of providing a polarized beam, which is important for 
nucleon spin physics, as documented in the EIC white
paper~\cite{Accardi:2012qut}. We briefly discuss the science opportunities 
enabled by the MuIC at BNL in this section. Quantitative studies will
be left for future work.

\subsection{Parton Distribution Functions}

Precision measurements of the structure functions at the MuIC in its new kinematic regime would enable a precise determination of the underlying parton distribution functions (PDFs) of the proton and other nuclei in a way complementary to hadron colliders, with a cleaner decoupling of the effects of QCD and quark flavor.  This would enable more precise cross section calculations to compare with measurements made at present and future hadron colliders like the LHC and the proposed FCC-hh, particularly for Higgs boson production where the gluon-gluon fusion process dominates and the cross section uncertainty from the PDFs$+\alpha_{\rm s}$ is 3.2\% at the LHC \cite{deFlorian:2016spz}. In particular, the MuIC, like the LHeC, would directly probe the PDFs at the scale and $x$-values for Higgs production at the hadron colliders. This would significantly reduce the cross section uncertainty obtained from global fits to previous DIS and pp collider data, to the level of 0.4\% for the gluon-gluon fusion process  and similarly for the $t\overline{t}H$ process \cite{Agostini:2020fmq}.  The MuIC also would further disentangle the flavor structure of the PDFs.

The measurements of the PDFs at the MuIC also would resolve some open issues in the proton content, as summarized in Ref.~\cite{Agostini:2020fmq}, such as the ratio of $u/d$ as $x \to 1$ and the universality of the light quark (and antiquark) sea. Additionally, fits to the structure function data allow for the precise measurement of $\alpha_{\rm s}$ and its running over a large $Q^2$ range.

\subsection{Nucleon Spin and 3-D Structure}

The nucleon spin is one of its fundamental properties.
It was found that quark polarization inside a nucleon only 
contributes to about 30\% of the total spin. Therefore, the 
majority rest of the nucleon spin must be carried by the 
gluon polarization and orbital motion of quarks and gluons.
To determine the contribution of gluon polarization,
a measurement of the helicity-dependent gluon distribution function, 
$\Delta g(x)$, especially in the small $x$ region, is crucial. 
Evidence for a finite gluon polarization in the proton for $x>0.05$ 
has been found by polarized pp collisions at
RHIC~\cite{Abdallah:2021aut}. 
However, the uncertainty on the overall gluon polarization 
is still rather large mainly because of the limitation in accessing 
the small $x$ region. For $x<0.01$  $\Delta g(x)$ is largely unconstrained. The EIC is projected to significantly improve the precision
of gluon polarization by accessing $x$ values down to 0.001 at 
$Q^{2} \approx 10$~GeV$^{2}$~\cite{Aschenauer:2012ve}, with an
integrated luminosity of 10~fb$^{-1}$. Assuming the same integrated
luminosity, the MuIC will extend the reach in $x$ down to $10^{-5}$, 
potentially providing a definitive answer to the gluon spin 
contribution. Furthermore, precise measurements 
of three-dimensional (3D) parton distribution functions, 
generalized parton distributions (GPDs) and
transverse-momentum-dependent (TMD) distributions, over
a wide range of $x$ values could provide a complete picture 
of orbital angular momentum of quarks and gluons inside 
the nucleon.

\subsection{Gluon Saturation at Extreme Parton Densities in proton and nucleus}

The gluon density inside the nucleon increases
dramatically toward small $x$ values. 
At extreme gluon densities, the nonlinear QCD
process of gluon-gluon fusion will start 
playing a key role to limit the divergence of
the gluon density. At a certain dynamic scale of
momentum transfer, known as $Q_{s}$, gluon splitting and fusion 
processes reach an equilibrium such that the gluon density 
is saturated, resulting in novel universal properties of 
hadronic matter. Examples of gluon saturation scales 
inferred from fits to HERA data (known as the GBW model)
\cite{GolecBiernat:1998js} are shown Fig.~\ref{fig1}, 
as discussed earlier, which is expected to be enhanced 
in a nucleus by a factor of $A^{1/3}$. The large $Q_{s}$ 
scale predicted at small $x$ values, especially in large 
nuclei, enables perturbative QCD calculations of nuclear
structure functions, as proposed in the color-glass
condensate (CGC) effective field theory~\cite{McLerran:1993ni}.
Predictions of signatures of gluon saturation in large nuclei
at EIC can be found in Ref.~\cite{Accardi:2012qut}.
As shown in the kinematic coverage of Fig.~\ref{fig1}, 
the EIC starts entering the domain of gluon saturation in 
gold nuclei at $x \approx 10^{-3}$, while the MuIC (and also LHeC) will
be well in the saturation regime for $x \approx 10^{-4}$--$10^{-5}$
at $Q_{s}$ of a few GeV. The MuIC will also probe the saturation
regime in the proton and other light nuclei for the first time,
which is not accessible by the EIC.

\subsection{Electroweak Physics}

The $Q^2$ reach of the MuIC, like the LHeC, would extend well above $M_{\rm W,Z}^2$, allowing studies of electroweak physics in the space-like regime. At such high $Q^2$ the charged-current cross section is comparable to the neutral current one, leading to many charged current interactions at high $Q^2$ and $x$.  One thing to note at the MuIC is that the final state leptons will be produced very forward, as discussed in Section~\ref{sec:kinematics}. However, the transverse momentum can still be sizable ({$\mathcal O$}(100) GeV) at high $Q^2>M_{\rm W}^2$, so even with final state neutrinos in charged-current interactions the kinematics can be measured from the final state hadrons. For smaller $Q^2$ this will be challenging experimentally, which also limits the applicability of charged-current events to study the flavor structure of the PDFs for $x \lesssim 0.01$. 

The MuIC also can be a facility to measure Higgs boson couplings, provided that the integrated luminosity is large enough given that the inclusive cross section is of the order 100~fb \cite{Agostini:2020fmq}. In particular, as Ref.~\cite{Agostini:2020fmq} suggests, a lepton-proton collider provides complementarity to the measurements made at pp colliders. For example, the Higgs boson decay to charm quarks is accessible at a lepton-hadron collider.

\subsection{Physics Beyond the Standard Model}

The MuIC will provide complementary capabilities of searching for
many physics phenomena beyond the standard model,
by providing a cleaner environment than $pp$ colliders. 
Particularly, it provides a unique opportunity to search for
new couplings to the second-generation leptons. For example,
searches for charged lepton flavor violation processes via
$\mu \to \tau$ at high energies can be carried out at the MuIC,
which complements the searches for $e \to \tau$ or $e \to \mu$ processes at HERA, EIC, and LHeC. Another example
is the search for leptoquarks, where the MuIC could be uniquely 
sensitive to leptoquarks that couple to muons and quarks (of any generation, as the lepton and quark generations do not need to match \cite{Diaz:2017lit}). In particular, anomalies observed in lepton universality in $B$ meson decays (e.g. \cite{LHCb:2021trn}) and in the measurement of the anomalous magnetic moment of the muon \cite{Muong-2:2021ojo} could be explained by leptoquarks \cite{Crivellin:2020tsz}.

Other search areas to which the MuIC could potentially improve sensitivity beyond the LHC are discussed in the LHeC report~\cite{Agostini:2020fmq}. This includes compressed supersymmetry scenarios, heavy neutrinos, dark photons, axion-like particles, excited fermions (particularly excited muons and excited muon neutrinos), and lepton-quark contact interactions. 

\begin{figure*}[t!]
\centering
\includegraphics[width=0.48\linewidth]{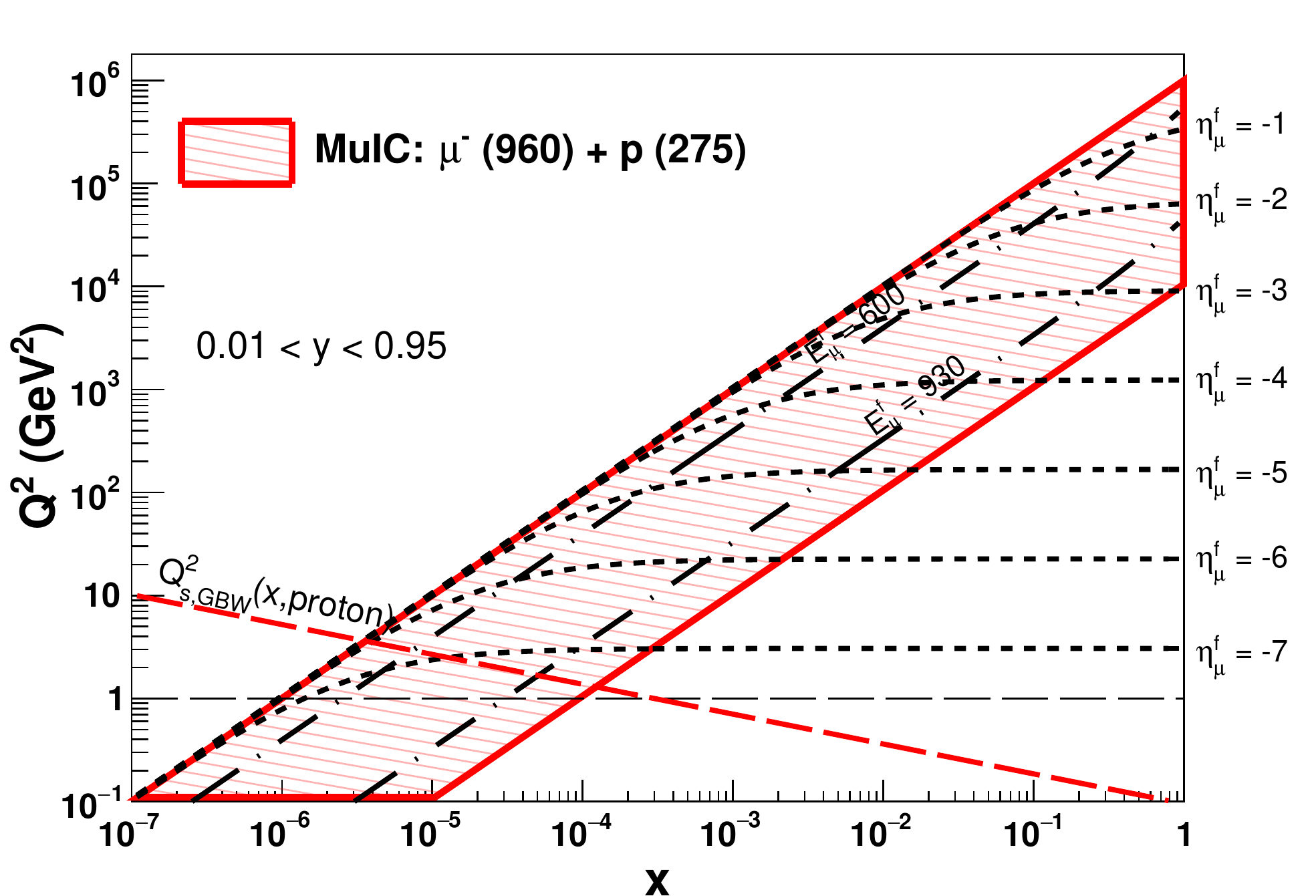}
\includegraphics[width=0.48\linewidth]{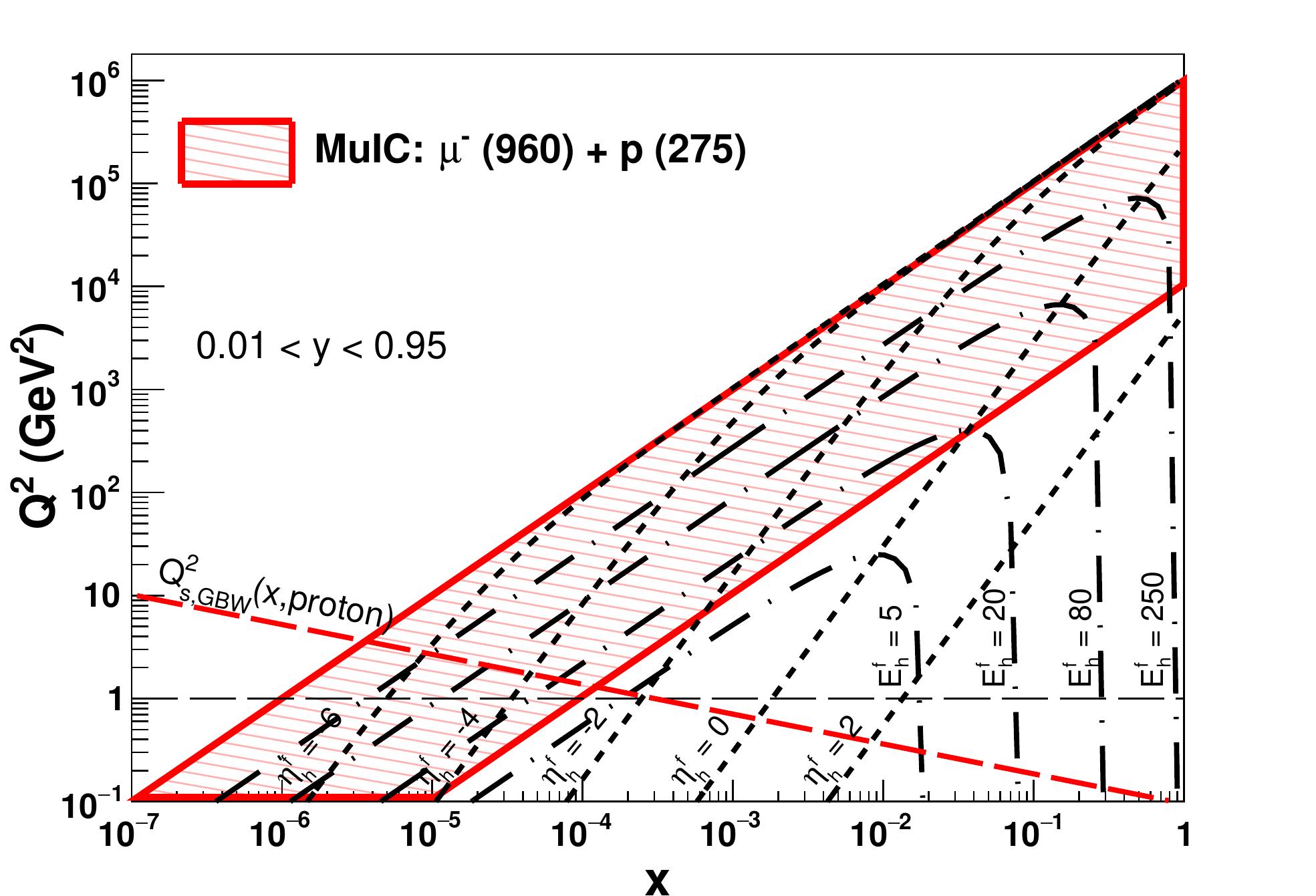}
\includegraphics[width=0.48\linewidth]{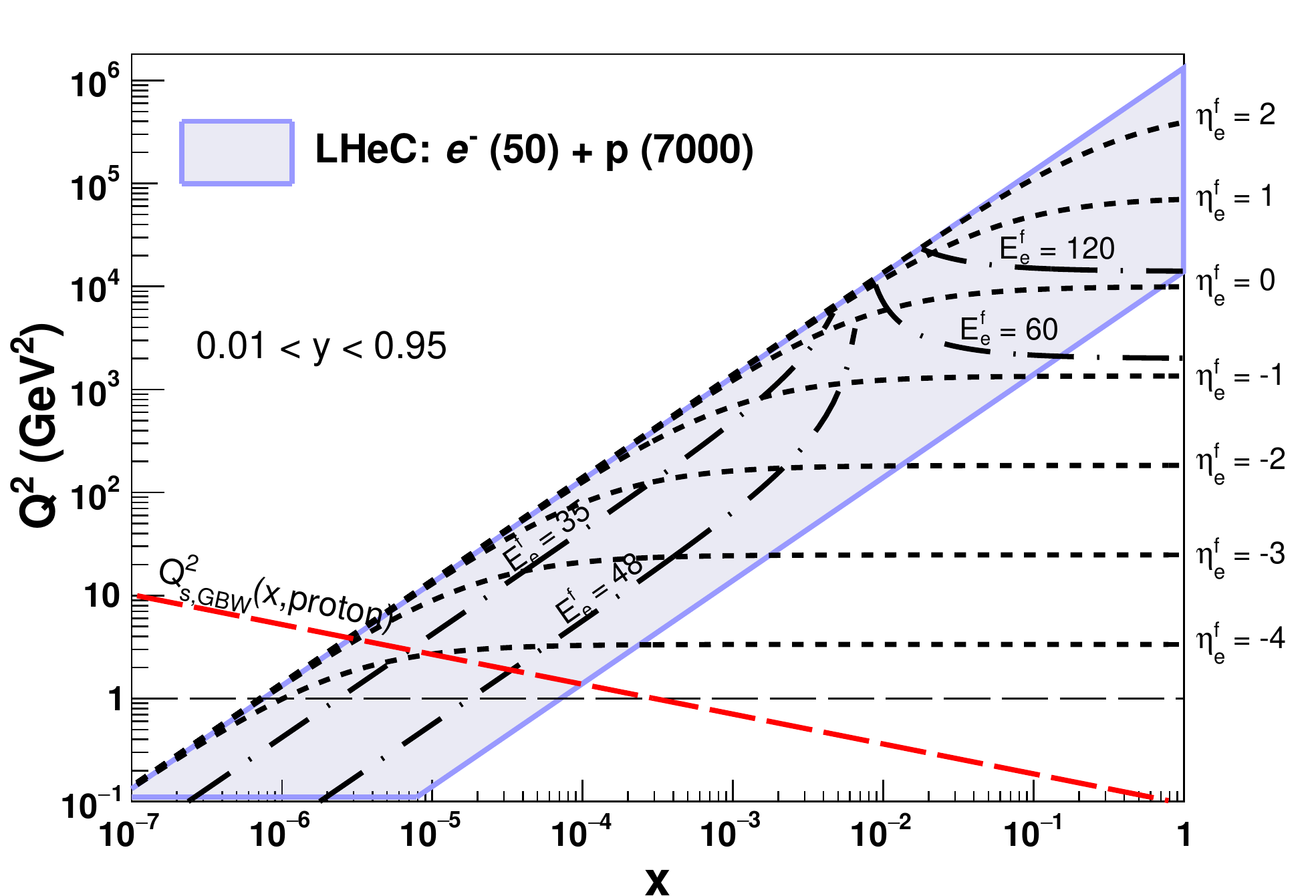}
\includegraphics[width=0.48\linewidth]{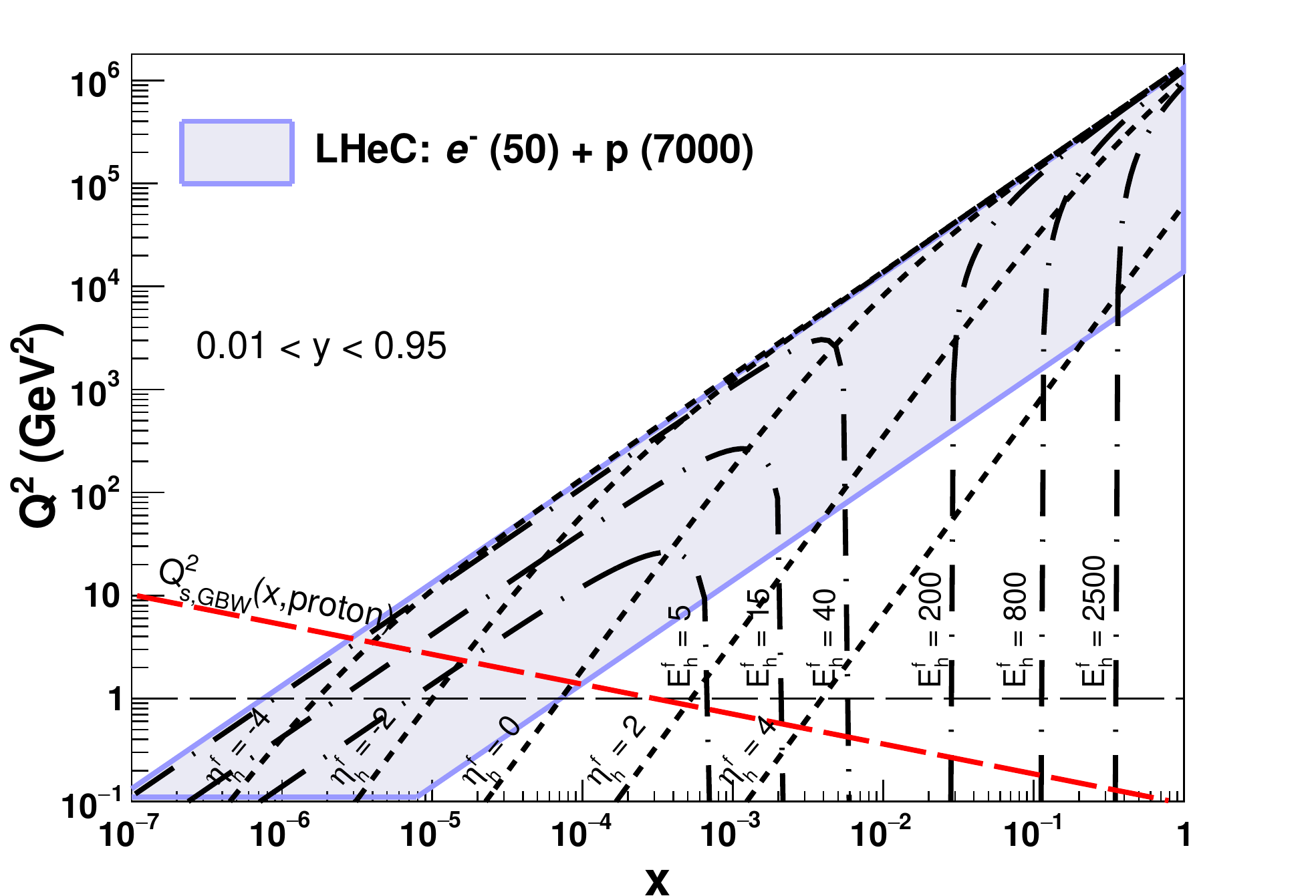}
\caption{Kinematics of scattered leptons (left) and final-state 
hadrons (right) in deep inelastic lepton-proton scatterings at 
the proposed MuIC at BNL (top) and LHeC at CERN (bottom) of the 
top energy. The dashed lines correspond to constant pseudorapidity 
(short-dashed) and energy (dot-dashed). The inelasticity ($y$) range 
is assumed to be $0.01<y<0.95$ (shaded areas).}
\label{fig2}
\end{figure*}

\section{Final-State Kinematics}
\label{sec:kinematics}

While the $Q^{2}$ and $x$ coverage determines the region of 
physics potential for a DIS facility, the final-state kinematics 
of the scattered lepton and produced hadrons are also important 
considerations. 
For perspective, in electron-proton collisions at the EIC (LHeC),
a relatively low energy 20~GeV (50~GeV) electron is scattered off 
a much higher energy 275~GeV (7~TeV) proton. In contrast, the MuIC would 
collide a nearly 1~TeV muon off of a 275~GeV proton, a more 
symmetric collision with much more momentum in the lepton direction.
This leads to complementary final state kinematics (not to mention 
a different lepton species), which poses different detector design
considerations. We focus here on a discussion of the kinematics and 
general requirements for a MuIC detector, but leave
the details of a detector design and simulation to future work.

Figure~\ref{fig2} shows the kinematics of scattered leptons
and partons (hadrons) in the $Q^{2}$ and $x$ map for the proposed 
MuIC (top) and the LHeC (bottom) at their highest energies 
for the inelasticity range $0.01<y<0.95$. We define the 
initial proton direction as the forward direction.
The dot-dashed lines indicate constant energy contours of the scattered lepton (left) and partons (hadrons, right). Likewise, 
the short-dashed lines indicate
constant pseudorapidity contours of the scattered lepton and
partons (hadrons),  as
determined by~\cite{Blumlein:1990dj}:

\begin{eqnarray}
    Q^{2}(x,\eta^{\rm f}_{\rm l}) &=& \frac{sx}{1+\frac{xE^{\rm i}_{\rm h}\exp(-2\eta^{\rm f}_{\rm l})}{E^{\rm i}_{\rm l}}}, \\
    Q^{2}(x,\eta^{\rm f}_{\rm h}) &=& \frac{sx}{1+\frac{E^{\rm i}_{\rm l}\exp(2\eta^{\rm f}_{\rm h})}{xE^{\rm i}_{\rm h}}}, \\
    Q^{2}(x,E^{\rm f}_{\rm l}) &=& \frac{1-E^{\rm f}_{\rm l}/E^{\rm i}_{\rm l}}{\frac{1}{sx}-\frac{1}{4(E^{\rm i}_{\rm l})^{2}}}, \\
    Q^{2}(x,E^{\rm f}_{\rm h}) &=& 
    \frac{sx(1-\frac{E^{\rm f}_{\rm h}}{xE^{\rm i}_{\rm h}})}{1-\frac{E^{\rm i}_{\rm l}}{xE^{\rm i}_{\rm h}}}.
\end{eqnarray}

\noindent where $E^{\rm i}_{\rm l,h}$ and $E^{\rm f}_{\rm l,h}$ are 
the energies of the lepton or hadrons (parton) before and after 
the scattering, respectively. The pseudorapidity of the final-state scattered 
lepton or hadrons (parton) is denoted by $\eta^{\rm f}_{\rm l,h}$.
Representative values of the scattered muon and 
hadron quantities for several benchmark points in $Q^2$ 
and $x$ are shown in Table~\ref{tab:benchmarks}: one at 
very high $Q^2$ and $x$, another at a medium $Q^2$ scale representative 
of Higgs boson production at the LHC, and a low $Q^2$ one near the 
expected gluon saturation scale.

\begin{table}[htp]
\caption{Benchmark kinematic points and scattered muon and hadron energies and pseudorapidities for a 960 GeV muon incident on a 275 GeV proton.
\label{tab:benchmarks}}
\begin{center}
\begin{tabular}{|l|r|l|c|r|c|r|}
\hline
 & $Q^2$ (GeV$^2$) & $x$ & $E^{\rm f}_\mu$ (GeV) & $\eta^{\rm f}_\mu$ & $E^{\rm f}_{\rm h}$ (GeV) & $\eta^{\rm f}_{\rm h}$ \\ \hline
High $Q^2$ & 500,000 & 0.5 & 180 & $0.5$ & 920 & $-2.4$ \\ \hline
Med $Q^2$ & 4000 & 0.01 & 600 & $-3.2$ & 370 & $-2.7$ \\ \hline
Low $Q^2$ & 3 & $5\times10^{-5}$ & 906 & $-7.0$ & 54 & $-4.2$ \\ \hline
\end{tabular}
\end{center}
\end{table}

As shown in Fig.~\ref{fig2} (top), the small-$x$ 
region at the MuIC corresponds to scattered muons in the very 
backward (lepton-going) direction. Reconstruction of the DIS
event kinematics would require the detection of muons at very 
small scattering angles from the beamline (down to a few mrad) 
as well measurements of momenta up to 1~TeV, both of which can 
be very challenging but not dissimilar to detecting final 
state muons in vector boson processes at a $\mu^+\mu^-$ collider. 
However, in DIS the kinematics are over 
constrained at leading order when one considers also the 
kinematics of the final state hadrons. In particular, 
one can avoid energy and momentum measurements entirely 
using the so-called ``double angle'' method using only the 
angles of the scattered lepton and hadrons~\cite{Bentvelsen:1992fu}, 
or rely entirely 
on the kinematics of the hadrons only, 
known as the ``Jacquet-Blondel'' method (JB). 
Moreover, unlike electrons, high energy muons will easily make it
out of the beam pipe and any surrounding shielding or magnets 
without much energy loss. Therefore, it is 
possible to set up muon detectors in the far backward direction 
outside the beam pipe. Here a beam crossing angle helps (e.g., 
25~mrad as for the EIC) to bring the most forward scattered 
muons at low $Q^2$ outside of the beam pipe in a longitudinal 
distance of meters rather than tens of meters. (An overall slight tilt of 
the muon storage ring also may help control the location and 
direction of where neutrino induced radiation emanates.)  In contrast, 
the high $Q^{2}$ region of the MuIC is more comfortably accessible 
near the central rapidity region for both scattered muons and 
final-state hadrons.

\section{Detector Requirements and Design Considerations}
\label{sec:detector}

Future extensive simulation studies of the detector will be required to fully
explore the feasibility and challenges of physics at the MuIC.
In this section, we discuss general detector requirements and
design considerations based on the final-state kinematics presented
in Section~\ref{sec:kinematics}, and present an estimate of
DIS kinematic variable resolutions based on feasible detector
technologies.

\begin{table}[b!]
\centering
\caption{List of assumed detector resolutions
for measuring various species of particles at MuIC.}
\vspace{0.5cm}
\includegraphics[width=\linewidth]{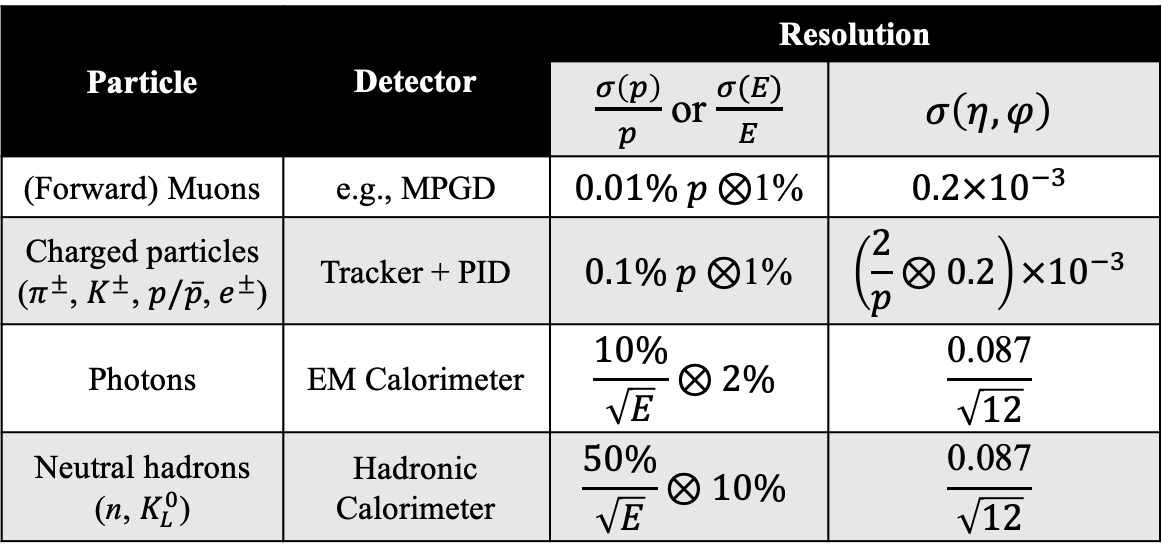}
\label{tab:table_resoltuion}
\end{table}

Reconstruction of the kinematic variables ($Q^{2}$, $x$ and $y$) with
good resolution is crucial to all DIS experiments. 
As pointed out earlier, the DIS kinematics
are over constrained and can be reconstructed from the scattered beam
lepton,  the hadronic recoil, or from a mixture of
both. Six methods of reconstructing $Q^{2}$, $x$ and $y$ are 
summarized in Ref.~\cite{Bentvelsen:1992fu}. 
To get an estimate of realistic resolutions at MuIC, we study 
the following three methods in this paper:
\begin{itemize}
\item Lepton-only method:
\begin{eqnarray}
  Q^{2}(E^{f}_{\mu},\theta) &=& 2 E^{i}_{\mu}E^{f}_{\mu}(1+\cos\theta), \\
  y(E^{f}_{\mu},\theta) &=& 1-\frac{E^{f}_{\mu}}{2E^{i}_{\mu}}(1-\cos\theta),
\end{eqnarray}
where $E^{i}_{\mu}$ and $E^{f}_{\mu}$ are incoming and scattered
muon energy, and $\theta$ is the polar angle of the scattered muon.
    
\item Jacquet-Blondel (JB) method based on hadronic activities:
\begin{eqnarray}
  Q^{2}(P,\gamma) &=& \frac{P^{2}\sin^{2}\gamma}{1-y(P,\gamma)}, \\
  y(P,\gamma) &=& \frac{F(1-\cos\gamma)}{2E^{i}_{\mu}},
\end{eqnarray}
where,
\begin{eqnarray}
P^{2} &=& (\Sigma_{h}P^{x}_{h})^{2}+(\Sigma_{h}P^{y}_{h})^{2}+(\Sigma_{h}P^{z}_{h})^{2}, \\
\cos\gamma &=& \frac{(\Sigma_{h}P^{x}_{h})^{2}+(\Sigma_{h}P^{y}_{h})^{2}-(\Sigma_{h}(E_{h}-P^{z}_{h}))^{2}}{(\Sigma_{h}P^{x}_{h})^{2}+(\Sigma_{h}P^{y}_{h})^{2}+(\Sigma_{h}(E_{h}-P^{z}_{h}))^{2}},
\end{eqnarray}
and ($E_{h}$,$P^{x}_{h}$,$P^{y}_{h}$,$P^{z}_{h}$) is a hadron four-momentum vector. 

\item Double Angle (DA) method based entirely on angles of scattered lepton ($\theta$)
and recoil hadrons ($\gamma$),
\begin{eqnarray}
  Q^{2}(\theta,\gamma) &=& 4(E^{i}_{\mu})^{2}\frac{\sin\gamma(1+\cos\theta)}{\sin\gamma+\sin\theta-\sin(\gamma+\theta)}, \\
  y(\theta,\gamma) &=& \frac{\sin\theta(1-\cos\gamma)}{\sin\gamma+\sin\theta-\sin(\gamma+\theta)},
\end{eqnarray}

\end{itemize}

\begin{figure*}[t!]
\centering
\includegraphics[width=\linewidth]{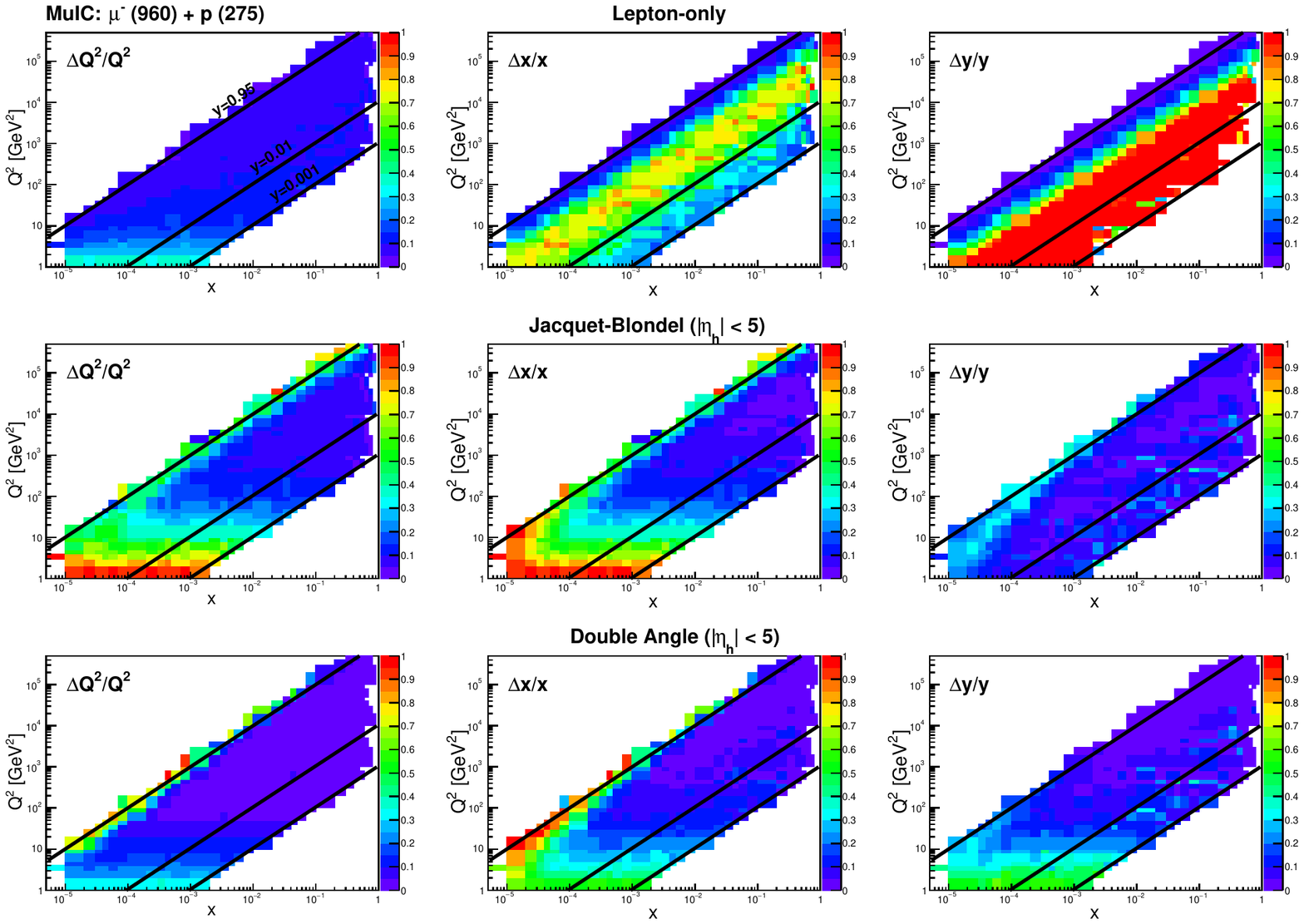}
\caption{Resolutions of $Q^{2}$, $x$ and $y$ as functions of 
$Q^{2}$ and $x$ in two dimensions
reconstructed with three methods: lepton-only (top), 
Jacquet-Blondel (JB) using hadronic activities within 
$|\eta_{h}|<5$ (middle), and Double Angle (DA) within 
$|\eta_{h}|<5$ (bottom). Lines indicated different values of $y$.
The assumed detector resolution parameters are summarized 
in Table~\ref{tab:table_resoltuion}.}
\label{fig3}
\end{figure*}

We make a set of assumptions on the detector resolutions
for measuring various species of particles at the MuIC,
listed in Table~\ref{tab:table_resoltuion}, and smear the
kinematic variables of particles generated by the PYTHIA 8 \cite{Sjostrand:2014zea} Monte Carlo event generator:
\begin{itemize}
    \item \textit{Scattered muons}: as mentioned 
    earlier, coverage of muon detectors to $\eta \approx -7$ in the far-backward region with excellent
    momentum and angular resolution is necessary. 
    We assume an angular resolution of 0.2~mrad, which should be well within the reach with the Micro Pattern Gas Detectors (MPGDs)~\cite{Pinto_2010} placed 10--20~m away from the interaction point. The momentum resolution will depend on details of the tracking system and magnetic field, so we simply assume a resolution factor similar to the performance of CMS muon systems~\cite{Sirunyan:2018fpa}, which is up to about 10\% at 1~TeV.
    
    \item \textit{Charged particles}: we assume that they will be measured by a high precision tracker (e.g., silicon sensors) with a resolution comparable to the CMS tracker~\cite{Chatrchyan:2014fea}. In addition, particle identification can be achieved by a time-of-fight (TOF) or a ring-imaging Cherenkov detector (RICH).
    
    \item \textit{Photons} are detected by electromagnetic calorimeters 
    (EMCal) with an energy resolution assumed to be similar to 
    that required for the EIC detectors~\cite{AbdulKhalek:2021gbh} 
    and an angular resolution similar to that of the CMS EMCal~\cite{Chatrchyan:2009qm}.
    
    \item \textit{Neutral hadrons}, such as neutrons and $K^{0}_{L}$, 
    are detected by hadronic calorimeters (HCal) with an energy
    resolution assumed to be similar to 
    that required for the EIC detectors~\cite{AbdulKhalek:2021gbh} 
    and angular resolution similar to that of the CMS 
    HCal~\cite{Collaboration_2010}. 
\end{itemize}

\begin{figure*}[t!]
\centering
\includegraphics[width=\linewidth]{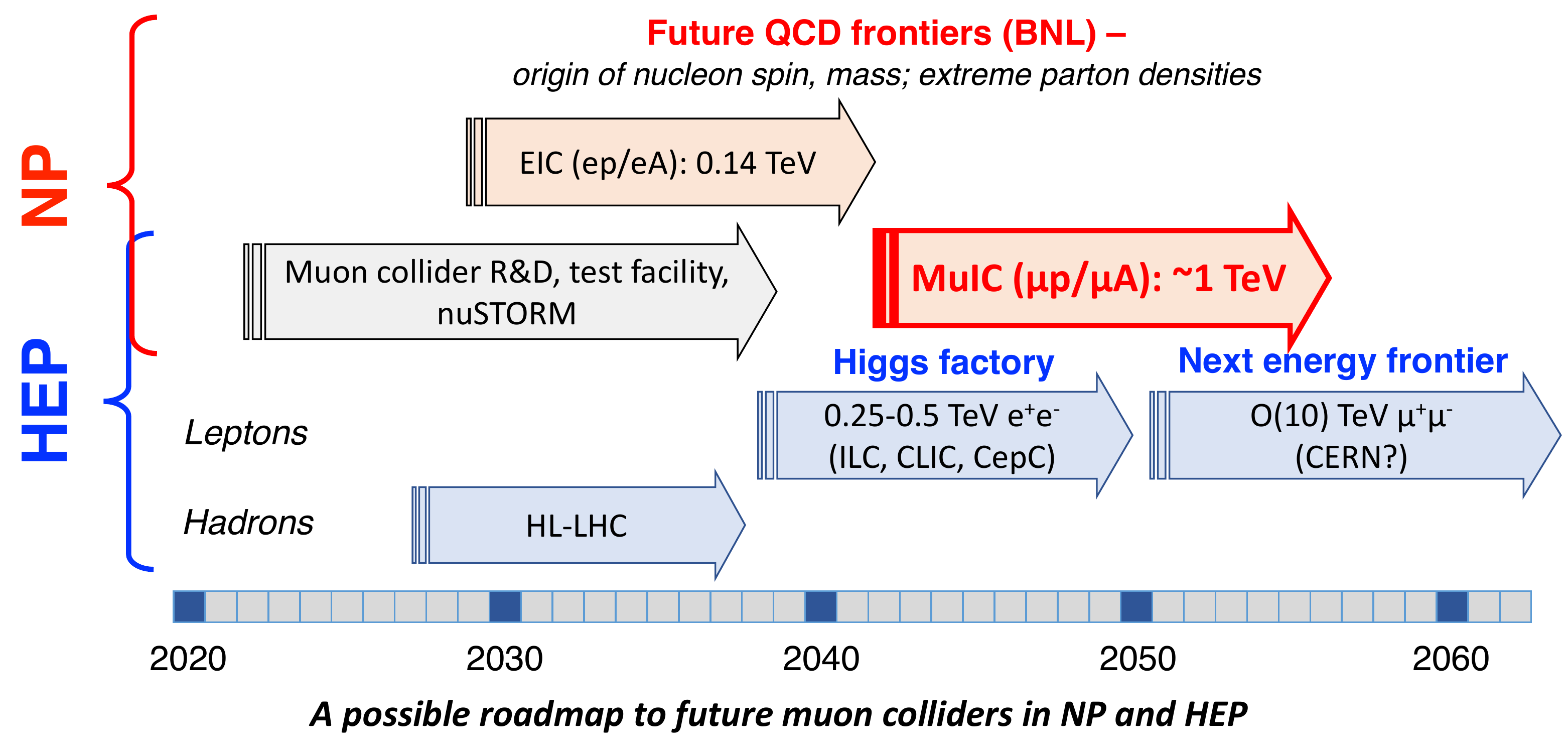}
\caption{A possible road map toward realizing future muon colliders with efforts of both the nuclear physics and high energy physics communities.}
\label{fig4}
\end{figure*}

The resolution of the reconstructed DIS kinematic 
variables, $Q^{2}$, 
$x$, and $y$ at the MuIC were studied using the 
PYTHIA 8 generator by smearing the final state particles by the  
resolution parameters listed in Table~\ref{tab:table_resoltuion}. 
The resulting resolutions (RMS) are 
presented in Fig.~\ref{fig3} in two dimensions as functions of
$Q^{2}$ and $x$ for the three reconstruction methods. 
Regions of different $y$ values are also indicated by lines for
$y$=0.95, $y$=0.01, and $y$=0.001.
A hadron acceptance of $|\eta^{f}_{h}|<5$ is assumed.

The lepton-only method gives good resolution of 10--20\%
in $Q^{2}$ over most of the kinematic region, 
but performs poorly in determining the $x$ and $y$ values, mainly
because of the very high muon momentum.
The DA method relying on exclusively the measurement of angles,
instead, provides much better resolutions of $x$ and $y$, 
except for $Q^{2}<10$~(GeV)$^{2}$. The JB method has excellent $y$ 
resolution down to very low $Q^{2}$ of 1~(GeV)$^{2}$ but 
underperforms in $Q^{2}$ and $x$ resolutions than the DA method 
for $Q^{2}<100$~(GeV)$^{2}$. The main limitation of JB and DA
methods in fact comes from the detector acceptance coverage, 
while the energy and angular resolutions are found to have 
negligible impact. At high $y$ regions (smaller $x$ and $Q^{2}$), 
the recoil hadrons start shifting 
toward very backward direction outside $|\eta_{h}|<5$. Therefore,
a wide coverage of detectors at MuIC, especially in the backward
direction for hadrons, is essential.
By optimizing the DIS kinematic reconstruction
based on measurements of both the scattered muon and hadronic 
activities, it should be feasible to achieve good resolutions 
at MuIC with conceivable detector technologies.

\section{A possible road map toward future muon colliders}
\label{sec:roadmap}

We present our view on a possible road 
map to a future combined facility for nuclear physics (NP) and high energy physics 
(HEP) based on muon collider technology in Fig.~\ref{fig4}. We assume that 
the muon collider community will need about 15 years to fully 
establish the feasibility of a muon collider via intense R\&D 
at a testing facility in order to be ready to construct a 
muon collider demonstrator. We propose that this effort be carried out jointly
between the NP and HEP communities, as this will attract more interest, 
resources, and be beneficial to both science programs.

In the USA, the EIC has been identified as the next QCD frontier 
by the NP community in the 2015 US Nuclear Physics Long Range
Plan~\cite{osti_1296778}. It has recently been endorsed by the US 
National Academy of Science~\cite{NAP25171} and the Department of 
Energy to be built at BNL by around 2029. The EIC is a unique machine, 
capable of colliding polarized electrons (up to 18~GeV) with 
polarized protons (up to 275~GeV) and a variety of nuclei 
with unprecedented high luminosity for the first time.
The success of the EIC is built upon its reuse of an existing RHIC 
that has been
in operation since 2000, in order to take advantage of the existing
tunnel, hadron beam, and infrastructure to minimize costs.
Following this approach,
constructing a MuIC after the EIC toward late 2040 by replacing 
the electron beam would open up new territory for QCD physics 
at a reasonable cost and share a strong synergy with the HEP 
community in advancing future collider technology.
Along the way, a nuSTORM-like facility may also be hosted at BNL
with stored muon beams.

At the high energy physics frontier, the upgrade of the High-Luminosity
LHC (HL-LHC) planned for 2027 aims to increase the delivered integrated luminosity
by a factor of at least 10 beyond the LHC's initial design,
bringing the promise of new discoveries in the next 10--15 years.
To plan for the next energy frontier era after the HL-LHC, there have been active discussions in the European
Strategy Planning Group, in the U.S. ``Snowmass'' community planning exercise, and 
in Asia (Japan, China). There appears to be a consensus that
a Higgs factory at an $e^+e^-$ collider (ILC, CLIC, CepC, FCC-ee) 
is a viable next step as a gateway to explore physics beyond the SM.
After the $e^+e^-$ Higgs factory, muon collider technology would
hopefully have matured with the MuIC as the cornerstone of a joint 
effort between the NP and HEP communities. Constructing a $\mu^{+}\mu^{-}$ collider 
at $\mathcal O(10)$~TeV would then open the next energy 
frontier to directly probe new physics, possibly re-using 
the existing facility and 
infrastructure at CERN, which significantly reduces the civil engineering cost.

In summary, a bright future of high energy nuclear and particle 
physics requires more collaboration to develop new, innovative 
ideas and technology. We lay out a possible path toward a future 
high energy muon collider with a proposed muon-ion collider at BNL 
as an intermediate step. This approach has high scientific and 
technological merits, and synergies interests of nuclear physics 
and particle physics communities. We look forward to developing 
more concrete steps toward an exciting future program.

This work is in part supported by the Department of Energy 
grant number DE-SC0005131 (WL), DE-SC0010266 (DA).

\bibliography{muic}{}
\bibliographystyle{apsrev4-1}
\appendix 

\end{document}